\documentclass[10pt,twocolumn,a4]{article} 
\usepackage{abstract}
\usepackage[utf8]{inputenc} 
\usepackage{amsxtra} 
\usepackage{amsthm}
\usepackage{amssymb}
\usepackage{amsfonts}
\usepackage{graphicx}
\usepackage{rotating}
\usepackage{color}
\usepackage{relsize}
\usepackage{epsfig}
\usepackage{subcaption} 
\usepackage{verbatim}
\usepackage[printonlyused]{acronym} 
\usepackage{setspace}
\usepackage{epstopdf}
\usepackage{authblk}
\usepackage{sectsty}
\usepackage[colorinlistoftodos]{todonotes}
\usepackage[colorlinks=true,allcolors=black]{hyperref}
\usepackage{textcomp,marvosym}

\usepackage{multirow}
\usepackage{graphicx}
\usepackage{epstopdf}
\pdfoutput=1

\sectionfont{\fontsize{12}{15}\selectfont} 
\usepackage{geometry} 
\definecolor{blue}{RGB}{0, 0, 150}
\definecolor{green}{RGB}{0,150,0}
\definecolor{red}{RGB}{200, 0, 0}
\definecolor{black}{RGB}{0, 0, 0}
     
\renewcommand{\l}{\left}
\renewcommand{\r}{\right}
\newcommand{\subeq}[2]{\begin{subequations}\label{#2}\begin{align}#1\end{align}\end{subequations}}

\newcommand{\eq}[2]{\begin{equation}#1\label{#2}\end{equation}}

\pdfoutput=1

\begin{document}

\title{Mechanisms of stochastic onset and termination of atrial fibrillation episodes: Insights using a cellular automaton model}
\author[1,\Yinyang]{Yen Ting Lin}
\author[2,\Yinyang]{Eugene TY Chang}
\author[3]{Julie Eatock}
\author[1]{Tobias Galla}
\author[2*]{Richard H Clayton}
\affil[1]{Theoretical Physics Division, School of Physics and Astronomy, The University of 
Manchester, UK}
\affil[2]{Department of Computer Science and INSIGNEO Institute for in-silico Medicine, The University of Sheffield, UK}
\affil[3]{Department of Computer Science, Brunel University London, UK}
\affil[ ]{{\normalfont\bf \Yinyang These authors contributed equally to this work.}}
\affil[ ]{{\normalfont\bf * r.h.clayton@sheffield.ac.uk}}

\date{\today}

\twocolumn[
  \begin{@twocolumnfalse}
	\maketitle
    \begin{abstract}

Mathematical models of cardiac electrical excitation are increasingly complex, with multiscale models seeking to represent and bridge physiological behaviours across temporal and spatial scales. The increasing complexity of these models makes it computationally expensive to both evaluate long term ($>60$ seconds) behaviour and determine sensitivity of model outputs to inputs. This is particularly relevant in models of atrial fibrillation (AF), where individual episodes last from seconds to days, and inter-episode waiting times can be minutes to months. Potential mechanisms of transition between sinus rhythm and AF have been identified but are not well understood, and it is difficult to simulate AF for long periods of time using state-of-the-art models.
In this study, we implemented a Moe-type cellular automaton (CA) on a novel, topologically correct surface geometry of the left atrium. We used the model to simulate stochastic initiation and spontaneous termination of AF, arising from bursts of spontaneous activation near pulmonary veins. The simplified representation of atrial electrical activity reduced computational cost, and so permitted us to investigate AF mechanisms in a probabilistic setting. We computed large numbers ($\sim 10^5$) of sample paths of the model, to infer stochastic initiation and termination rates of AF episodes using different model parameters. By generating statistical distributions of model outputs, we demonstrated how to propagate uncertainties of inputs within our microscopic level model up to a macroscopic level. Lastly, we investigated spontaneous termination in the model and found a complex dependence on its past AF trajectory, the mechanism of which merits future investigation. 

	\end{abstract}
\end{@twocolumnfalse}
\vspace{10pt}
]

\section*{Introduction}
Mathematical and computational models have become an increasingly popular tool for investigating biological and physiological systems. The quantitative capabilities of models can provide both unique insights into the mechanism of a problem and predictive power beyond experimental or clinical preparations. Once developed, a model can be used to test and generate future hypotheses in a way that may not be possible in experimental settings. The holy grail of computational biology is to develop a comprehensive model which describes both mechanistic properties---for example, the detailed molecular dynamics of biochemical interactions in a living organism---and subsequent emergent phenomenon.  

These comprehensive models are largely constrained by current computational power.
Instead of comprehensive models, a more adaptable approach is to select a relevant spatial and temporal scale of the phenomenon and devise models which suit a particular research question.  Thus, for the same biological or physiological systems, a wide spectrum of models may co-exist, which aim to explain and predict the physiological process at different length or time scales. By analysing these models separately, researchers can gain a deeper understanding in the regimes when the different quantitative models are adequate. Bridging these models provides a way to propagate results derived from one model into inputs for another model.

In computational cardiac electrophysiology, there exist a range of models, which have been used to examine how sub-cellular electrical processes influence the diffusion of activation wavefronts across the heart \cite{Trayanova:2014}. Computing detailed biophysical models involves solving large systems of coupled ordinary or partial differential equations, which is computationally demanding. This limits both the number of simulations that can be run as well as their duration. It is therefore difficult to explore the sensitivities of a given model to input parameters and initial conditions. This in turn means that model results cannot easily be translated into inputs for models at other scales, such as those describing progression of patients through care pathways \cite{Lord:2013}. 

Atrial fibrillation (AF) is a cardiac arrhythmia that remains poorly understood despite progress in the development of detailed cardiac electrical models, experimental work and clinical studies. AF presents a prevalent heart rhythm disorder which significantly increases stroke risk \cite{Camm:2010}. Improving identification, management and treatment of AF remains an important challenge \cite{Nattel:2002}. AF consists of episodes of rapid and self-sustaining electrical excitation in the atrium of the heart, which punctuate periods of normal sinus rhythm when activation is driven by the heart's natural pacemaker. As the disease develops, episodes of AF become longer and more frequent until AF becomes permanent. Episode duration can vary between seconds and weeks, constitutes the basic clinical marker to classify AF progression in patients. The ability to model and predict episode duration for a given patient would therefore be of significant clinical interest. In a previous publication \cite{Chang:2016} we described a biophysically motivated agent-based stochastic model to simulate progression of AF in a patient from first diagnosis, based on generating a time series of AF episodes with varying durations. The model parameters, which predicted episode start times and durations, were estimated from the literature where possible. 

The duration of an AF episode may best be analysed by studying mechanisms underlying its initiation and termination. AF is thought to be driven by rapid and self-sustaining electrical activity predominantly occurring in the left atrium \cite{Nattel:2002,Haissaguerre:1998}. Re-entry, in which a circulating activation wave continually propagates into recovering tissue, often sustains AF. Several mechanisms have been associated with re-entry initiation, including atrial fibrosis \cite{Schotten:2011}, pulmonary vein triggers \cite{Haissaguerre:1998}, and action potential and conduction velocity restitution \cite{Qu:1999}. Meanwhile, mechanisms of AF termination remain a poorly researched area, in part due to the computational cost of evaluating complex cardiac electrophysiology models. Initiating and maintaining an AF episode up to its termination in a simulation of a biophysically detailed model requires significant computational resources \cite{Liberos:2016}, in particular if the episode lasts for more than a few seconds and/or a detailed atrial geometry is used.

Cellular automata (CA) models of the electrical activity on the surface of the heart are a simplified representation of cardiac electrophysiology, and were used in the very first simulation of AF \cite{Moe:1964}. Since the original five-state Moe description \cite{Moe:1964}, more complex models have been devised \cite{Bub:2002,Manani:2016}. They provide an intuitive way of describing how cardiac cells activate (depolarise) and deactivate (repolarise) by using simple update rules for the state of a single cell. These are usually based on the present states of the cell itself, and of its nearest neighbours. CA models are simple to program and computationally cheap to run, allowing large numbers of simulations for little cost. 

The motivation of this study was to adopt a CA model as a computational platform to investigate stochastic initiation and termination of AF episodes. Our contribution can broadly be summarised as follows. First, in contrast to previous cellular automata models \cite{Bub:2002,Manani:2016}, which used a simplified geometry, we generalised to a more realistic topology representing the left atrium. Whilst the geometry is still stylised we think this is a step towards reality. We show that the model is capable of inducing and terminating AF episodes stochastically. These phenomena are in line with the predictions of current state-of-art mechanistic models, and we are confident the CA model captures the essential dynamics of the real physiological system. Thus, we propose that CA models are a reasonable compromise between reality and computational efficiency when large numbers of simulations are required. Second, we present a framework to analyse and infer the rate of stochastic initiation and termination of AF episodes. With the ability to run large numbers of simulations, we were able to accurately quantify these rates. This is necessary in order to be able to predict -- in a statistical sense -- the future progression of patients at a longer time scale.
For example, these rates can be used to connect the CA model to the model we proposed to represent AF progression over years and decades \cite{Chang:2016}. We propose a framework of statistical analysis of patient trajectories, and apply it to a set of patient trajectories, generated from the CA model. We believe the ideas suggested may also be applicable to data from mechanistic models of other physiological systems, when computational resources are available to generate sufficiently many sample paths from such models.

\section*{Methods}
 
\subsection*{Model geometry}
Electrical activation was modelled on an idealised spherical geometry, representing the left atrium of the human heart. We did not include the right atrium because the main drivers of AF are believed to originate in the left atrium.
The volume of the sphere representing the left atrium was set to 40~mL \cite{Lang:2006}, corresponding to a radius of $21.2$~mm, to which the geometry was rescaled, to obtain a unit sphere.  

\begin{figure*}[ht]
\begin{center}
\includegraphics[width=0.75\textwidth]{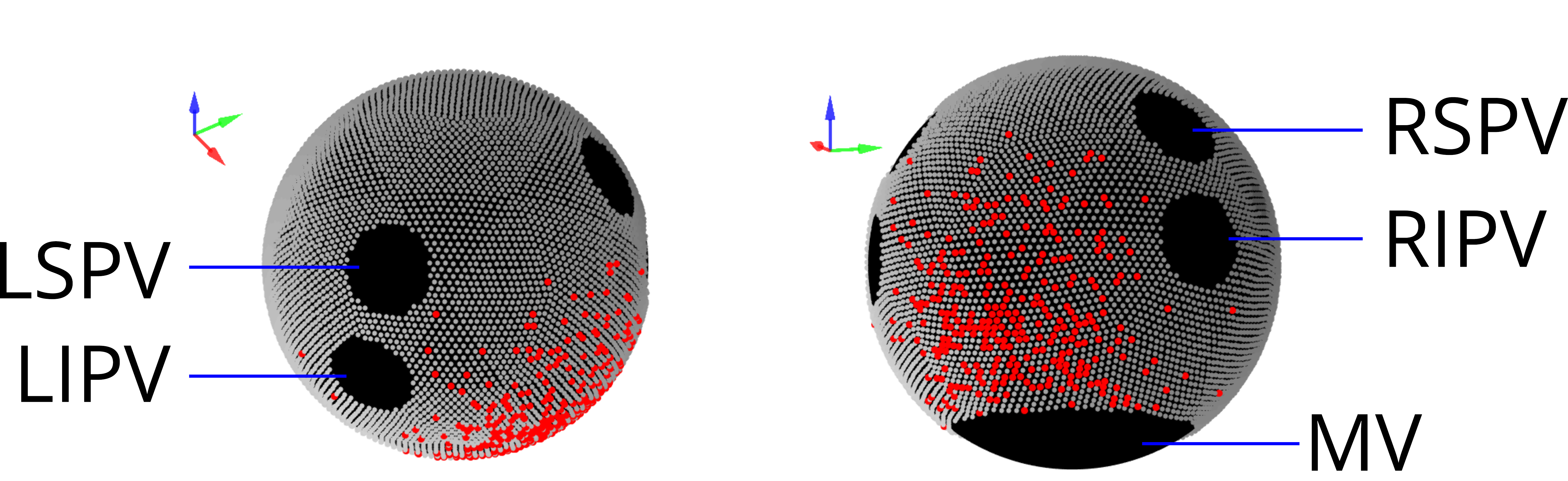}
\caption{\label{fig:Fig1}Visualisation of the spherical geometry representing the left atrium, with nodes distributed regularly over the surface. Anatomical features (in black) were rendered electrically inactive. LS/LIPV: Left Superior/Left Inferior Pulmonary Vein. RS/RIPV: Right Superior/Right Inferior Pulmonary Vein. MV: Mitral Valve. Fibrotic cells (in red) were distributed randomly over a disk centred on the posterior atrial wall.}
\end{center}
\end{figure*}

We implemented a Moe type cellular automaton \cite{Moe:1964} in which the dynamics take place on discrete nodes on the surface of the sphere.
To place the nodes on the spherical surface as uniformly as possible, we used an icosahedral dissection \cite{Saff:1997} to distribute 10242 points evenly on the sphere, as visualised in Figure \ref{fig:Fig1}. We also investigated an alternative way to distribute nodes using an Archimedian spiral \cite{Huettig:2008}; this method can be generalised to non-spherical surfaces.  

Polar coordinates $\l(\theta,\phi\r)$ were used to specify the locations of the nodes. We defined the anterior and posterior direction to be $\l(\theta,\phi\r)=\l(\pi/2, 0\r)$ and $\l(\pi/2, \pi\r)$ respectively.  

On the geometry, the anatomical objects---four pulmonary veins (PVs) and the mitral valve (MV)---were set to be electrically inactive. The mitral valve, modelled as a circular area centred on the south pole $\l(\theta,\phi\r)=(\pi,0)$, was estimated to have circumference 85mm \cite{Otto:2009}. The four PVs were modelled as circular areas with a base radius 5mm \cite{Kim:2005} corresponding to $0.236$ scaled units. The PVs were placed at $\l(\theta,\phi\r)= \l(2\pi/5\pm \pi/10, \pm \pi/3\r)$. Nodes in these areas were permanently removed, and the remaining nodes comprised the substrate for the cellular automaton to take place.

\noindent {\bf Fibrosis.} Fibrosis on the posterior atrial wall is thought to play an important role of inducing AF re-entry \cite{McDowell:2013,Schotten:2011,Trayanova:2014}, and was modelled by removing nodes in the corresponding area.
To model the spatial heterogeneity of fibrosis, we removed nodes according to a probability distribution set to be normally distributed, centred at $\l(\theta,\phi\r)=\l(0.65 \pi, 0\r)$, with a standard deviation equal to $0.4$ sphere radii.
The number of nodes removed (denoted FI) quantified the severity of fibrosis. Time dependent fibrosis was not investigated in the current study, as structural modelling of atrial tissue with fibrosis occurs at a time scale much slower than that of re-entrant activity \cite{Burstein:2008}.

\subsection*{Dynamics of the CA model}
A multi-state Moe-type cellular automaton was used to represent electrical excitation in each node (or `cell'). Each node on the sphere could be in one of a number of discrete states, labelled $0,1,2,\dots$. In this type of discrete-time model, an action potential is represented by a time-delay, during which an excited cell may trigger neighbouring cells within an interaction radius but cannot itself be re-excited. In our model, the cell was deemed `at rest' at state 0 and `activated' if its state was greater than $0$. A cell at rest would become excited if the number of `recently-excited' neighbours in a local radius exceeded a threshold, upon which it changes from state $0$ to state $RP$, the Refractory Period or action potential duration, at the next time step. A neighbour was considered to be `recently-excited' if it had been activated in the past $4$ time step. This number was chosen to achieve realistic spread of excitation (see below). 
Following excitation, the activated cells reduced their state by one each time step until the state reached 0, i.e.,  the `rest' state. Each discrete time step in our simulation corresponds to approximately $2.5$ms in real time. In sinus rhythm $RP$ took values of about $120$ in the model (variations are described below), this represents a physiological refractory period of $300$ms.

To avoid grid discretisation effects on the simulations due to non-uniformities of the icosahedral mesh, the interaction radius between cells on the sphere was set to be greater than the length scale of the typical inter-node spacing (for complications, see Ventrella \cite{Ventrella:2011}). 
The speed at which an excitation wavefront could propagate (conduction velocity) was determined by two free parameters: the search radius and the threshold of number of active neighbours. We carefully calibrated both the active neighbour thresholds ($=8$ nodes) and local search radius (2.544 mm), corresponding to a region containing $\approx 36$ nodes to achieve a baseline conduction velocity across the sphere of $0.5$ m/s. Thus, the total time taken to travel across the unit sphere (defect-free) from north pole to south pole was $\approx133$ ms.

\noindent {\bf Sinus rhythm.} The sinus node is located in the right atrium, so in our model sinus rhythm was represented by the regular activation of a region of cells proximal to the right pulmonary veins (a circular area centred at $\l(\theta,\phi\r) = \l(5\pi/12, \pi/2\r)$ with radius $1.696$ mm), which is typically the site of earliest activation in the left atrium following right atrial activation. The sinus period was set at 1Hz for all simulations, unless otherwise specified.
 
\noindent {\bf Pulmonary vein triggers.} Bursts of spontaneous activation near the PVs are thought to be triggers of re-entry \cite{Haissaguerre:1998}. To model PV bursts, a 2mm annulus around each of the four PVs was set to be capable of auto-excitation. In each time step with probability $p$, one node in these annulus regions and its surrounding nodes (set as those points within  $2.12$ mm to the selected node) spontaneously fired to its maximal state. The location of this spontaneous firing was chosen uniformly on the annuli. The probability $p$ quantifies how often these bursts occur; the corresponding burst rate in a continuous time setting can be computed using $p/$time step$=$ continuous-time bursting rate $BR$, which is set as a model parameter. Note that triggers were stochastic and occurred \emph{on average} $BR$ times per second, rather than periodically every $1/BR$ seconds.

\noindent {\bf Restitution.} To model the effect of \emph{restitution} where the refractory period (RP) (i.e. action potential duration) of a cell shows sensitivity to its previous rate of excitation, we implemented the following (non-dimensionalised) formula \cite{Kalb:2004}
\begin{equation}
\rm{RP} = \lfloor 121 \times \left[1-B\exp(-\rm{DI}/K)\right]\rfloor, \label{eq:restitution}
\end{equation}
where DI is the diastolic interval (the quiescent interval between the end of one activation and the following beat), and $B$, $K$ are parameters controlling the steepness of the curve. $K$ had units of discrete time ($=2.5$ ms), and $B$ was dimensionless. The floor function $\lfloor \rfloor$ enforced that RP was an integer, which in combination with scale factor 121, allows a maximal RP of 120.
RP was subject to a minimum of $64$ time units, i.e. $RP = \max(Eq. (\ref{eq:restitution}),64)$. This equates to $160$ ms (considered the shortest physiologically relevant RP).  
We investigated the dependence of the transition rate into AF episodes on parameters $B$ and $K$.

\subsection*{Implementation}
The model was implemented with custom code written in C++, and is publicly available on Github at \href{https://github.com/dblueeye/atrial-fibrillation-cellular-automata}{{\bf https://github.com/dblueeye/atrial-fibrillation-cellular-automata}}. Links to sample movies may also be found. The simulation ran at 16x speedup, i.e. 16s of simulated time could be evaluated in 1s. The skeleton code of the simulation is detailed below to clarify implementation steps:
\begin{enumerate}
\item Initiation: Set up location of the nodes on the sphere. Remove nodes on areas of the PV and MV. For each sample run, model fibrosis by removing a fixed number of nodes according to a spatial probability distribution. Briefly, assign a probability to each node, generated from a normal distribution centred at $\l(\theta, \phi\r) = (0.65 \pi, 0)$ with a standard deviation $8.48$. Then, arrange the probabilities into a list and compute the cumulative probability distribution $F(i)$ with respect to the list. The inverse transform sampling was applied to the discrete distribution to select the node to be taken out. Repeat the procedure until FI numbers of nodes were taken out.
Generate and store the nodes representing sinus node breakthrough. Generate and store a list of possible PV bursting locations and the nodes which would burst in a group. Generate a neighbourhood map between the nodes. 
\item Sinus node breakthrough (SN): Check if in this time step SN breakthrough occurs. If so, activate the nodes of SN to their maximal state as follows: Using the cycle length (CL) (time between SN pacing)  and refractory period (RP) from the previous cycle, compute DI$=$CL$-$RP. Use Eq.~\eqref{eq:restitution} to compute and update the refractory period (RP) of this node, and activate its state to RP.
If in this time step SN breakthrough does not occur, the state of SN breakthrough nodes is reduced by 1. 
\item PV bursts: With probability $p$ there will be a PV burst. If this happens choose one of the locations where PV bursts can take place. As described above a group of nodes in that region is activated to their maximal state, and the new RP is computed and updated using Eq.~\eqref{eq:restitution}.  
\item Rest of the nodes: For the remaining nodes, check if any neighbours in the interaction range have been activated in the past 4 time steps ($10$ ms). If so, this node is activated to its maximal state, RP is again updated according to Eq.~\eqref{eq:restitution}. Otherwise, the state of the node is reduced by 1.
\item Repeat from 1 until end of simulation. 
\end{enumerate}

\begin{figure}[t]
\begin{center}
\includegraphics[width=0.48\textwidth]{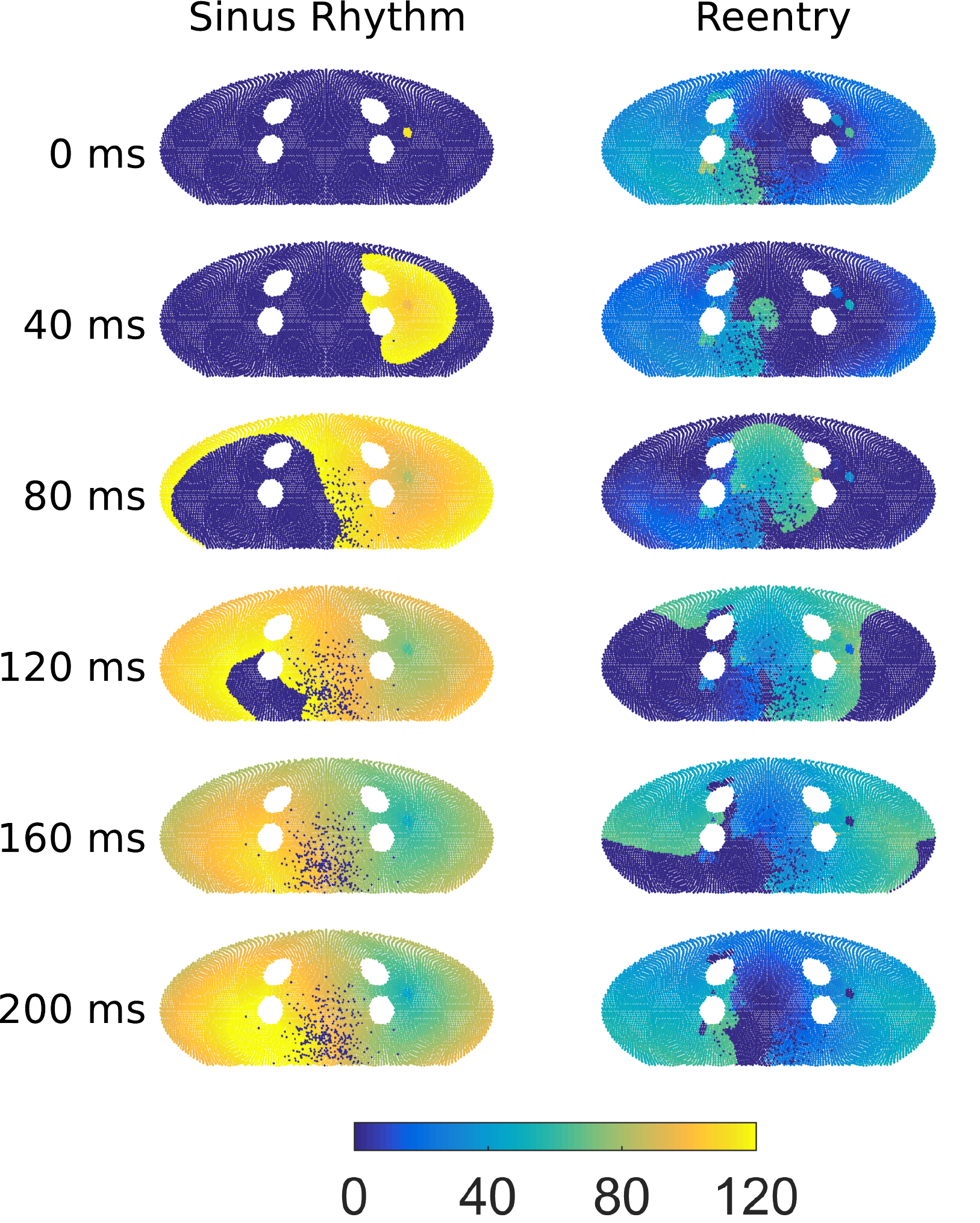}
\caption{\label{fig:Fig2}Snapshots of the simulation using a Mollweide projection. The centre of the diagram is at the posterior atrial wall. Fibrotic nodes (red), which cannot be initiated, are set to a constant state. In sinus rhythm ({\bf left panel}), sinus node breakthrough starts proximal to the right pulmonary veins (PVs), and cells are immediately excited from 0 to (maximal refractory period) state 120, decreasing its state by 1 each time step until it reaches 0. Cells nearby are excited to 120 if the number of neighbouring cells which are excited exceeds 8, and this starts a wavefront of activation over the sphere, with slower activation through fibrotic areas and around PVs. In re-entry ({\bf right panel}), existing wavefronts self perpetuate across the domain, and sinus node breakthrough does not initiate wavefronts of excitation. Refractory period restitution has meant that cells excite to a lower state compared to sinus rhythm, and this also leads to a shorter wavetail.}
\end{center}
\end{figure}

\section*{Results}
\noindent {\bf The CA model exhibits probabilistic initiation of AF.} 
Simulation results were visualised using an equal-area Mollweide projection \cite{Snyder:1997}, shown in Fig.~\ref{fig:Fig2}. During sinus rhythm without PV bursts (left panel), activation of the left atrium began by sinus node breakthrough near the right PVs; wavefronts passed around the larger PVs smoothly with a conduction speed of 0.5m/s. When wavefronts passed through areas of fibrosis, conduction slowing and conduction block were observed occasionally when the number of activated nearest neighbours remained sub-threshold. When PV bursts were introduced, triggers initiated activation near single PVs at a constant rate; in some simulations this led to transient re-entrant wavefronts forming, and in certain cases these became permanent re-entrant wavefronts (right panel). Movies have been  uploaded to Youtube, and can be found on the Github project page, see Supporting Information.

Whilst the complete course of the stochastic process (for each node) could be stored, the resulting data file would be impractically large. Instead, we evolved the CA without exporting the dynamic states at all time steps. As our aim was to investigate statistical properties of the system initiating and terminating AF (defined as self-sustained activity differing from sinus rhythm), two AF classifiers were developed. We stored only the seeds of the pseudo-random number generator of those sample paths, which were classified as `in AF' (details are described below). If needed, the collected seeds could recreate the sample paths for subsequent analyses. This procedure permitted generation and storage of up to $10^4-10^6$ simulation runs; necessary to accurately compute the statistics of AF episodes, including the sampling of rare events. 

\begin{figure*}[t]
\centering
 \includegraphics[angle=270,width=0.7\textwidth]{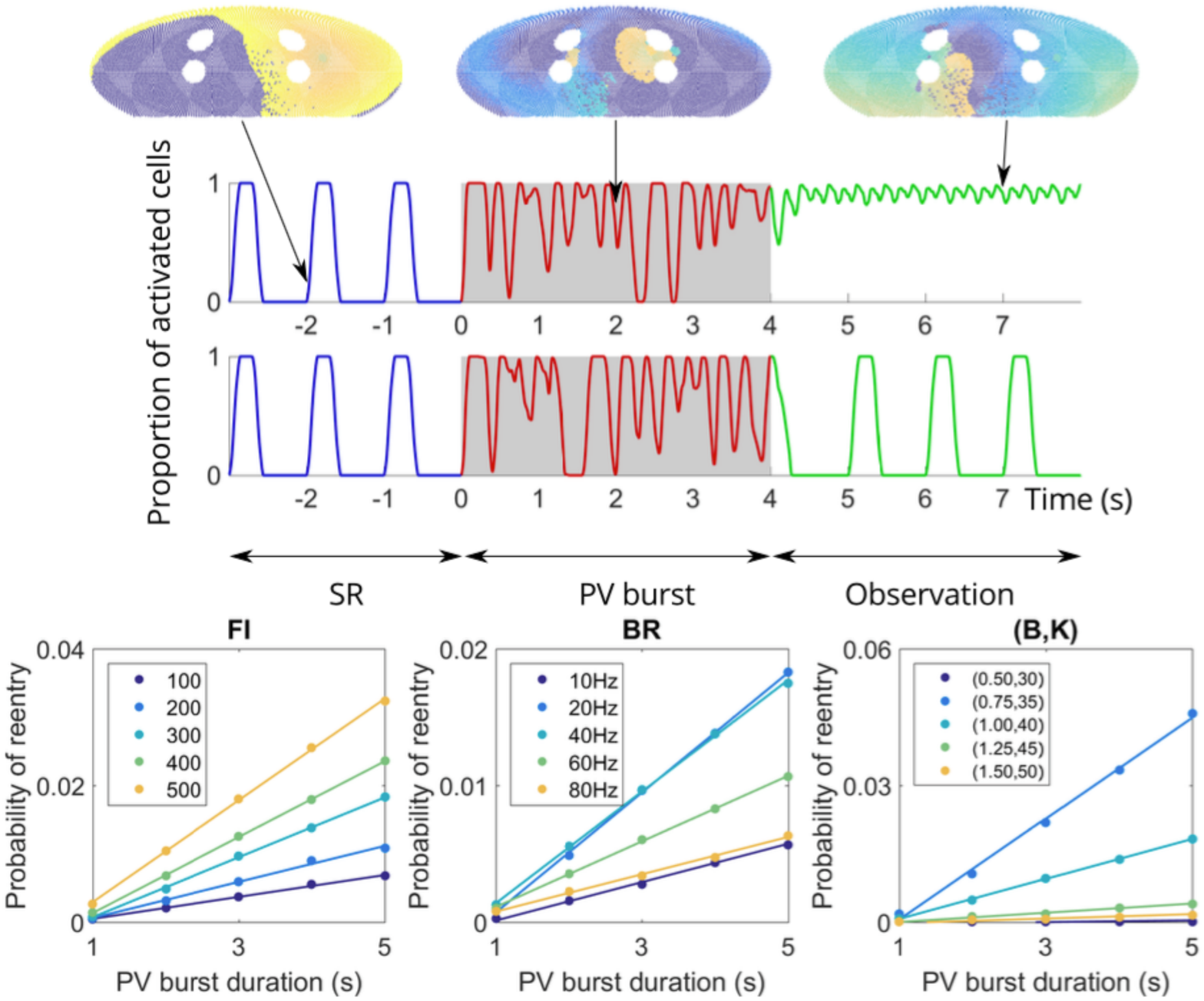}
  \caption{\label{fig:Fig3}{\bf Top:} Protocol for investigating AF initiation. Starting in sinus rhythm (SR), PV bursts of up to 5s were initiated, after which a 10 second observation window with sinus pacing was simulated to probe existence of AF (top/bottom sample path: with/without AF. In the top sample path, the sinus breakthrough region cannot be excited by sinus pacing because re-entrant waves keep re-exciting the region from state 0.) {\bf Bottom:} Probability of re-entry (the proportion of simulations finishing in re-entry over $10^5$ sample paths ), as a function of the model parameters fibrosis density ($FI$), PV bursting rate ($BR$) and restitution steepness $(B,K)$ as well as PV burst duration. The slope of the curves quantifies the continuous-time rate to induce AF re-entry. Discrete markers: simulation results; continuous lines: best linear fits.}
\end{figure*}

\noindent {\bf The probability of initiating spontaneous AF re-entry depends non-trivially on the model parameters.} 
An exploration of the model parameter space was undertaken to determine the primary mechanisms of re-entry initiation. Each simulation was started in sinus rhythm (by setting $p=0$), then PV bursts of varying time durations were initiated by setting $p\equiv BR/400$, to simulate PV triggers on the domain. 
Following a period of time with PV bursts, $p$ was reset to 0, and the simulation evolved for a further 10s observation window (Fig. \ref{fig:Fig3}, snapshots). The proportion of activated cells (in all nodes excluding fibrotic ones) at each time point was tracked, as a simple classifier of re-entry; simulations in which the proportion of activated cells remained non-zero over the entire observation window were deemed in re-entry. An example can be found in the top panel of Fig. \ref{fig:Fig3}, the first time series remains in re-entry, whereas the second time series returns to sinus rhythm. The proportion of sample paths leading to re-entrant wavefronts determined the probability of a given parameter set causing re-entry.

We varied the following parameters: number of fibrotic cells ($FI$), PV bursting rate ($BR$), and restitution steepness ($B,K$). Baseline simulation parameters were: $FI = 300$ (points), $BR = 20$ (Hz), $B = 1.0$ and $K = 40$ (discrete-time unit, $=100$ ms), and we point-mutated the parameters ($FI,BR,B,K$)---see Tab.~\ref{tab:parameterset} for the full range. 
For each parameter set, PV burst duration was varied from $1-5$~s, and $10^5$ sample paths were generated to compute the probability of inducing re-entry. Results are summarised in the lower panels in Fig.~\ref{fig:Fig3}.

We found the probability of AF re-entry depended linearly on the duration of the PV bursting when this duration was $\leq5$ seconds. This suggests AF initiation may be modelled by a simple coarse-grained model in continuous time, in which initiation occurs with constant rate, written as follows:
\eq{
\text{Sinus Rhythm} \longrightarrow \text{AF} \quad \text{with a rate } r_1,
}{eq:onlyActivationReaction}
where the transition rate $r_1$ is the slope of the linear response shown in the bottom left panel of Fig.~\ref{fig:Fig3}. We performed a linear fit to the numerical data, and found the rate was monotonically dependent on fibrosis: $r_1$ is larger for higher $FI$. Estimated values for $r_1$ are reported in Table \ref{tab:parameterset}. 

We found a non-monotonic relation between re-entry probability and PV burst rate $BR$, seen in the bottom middle panel of Fig.~\ref{fig:Fig3}. A PV burst was most likely to induce AF when $BR$ was between $20$ and $40$ Hz. Similarly, the rate into re-entry had a non-monotonic response to the restitution parameters. These observations suggest that the CA model is able to capture complex interplay between the mechanisms inducing AF.
\noindent{\bf Estimating time of AF initiation using a dynamic classifier. }
In the previous section, we investigated the hypothetical case where we controlled PV burst duration independently and subsequently observed for AF. In reality, PV bursts occur at random and cannot be simply turned off physiologically - AF may have initiated before the end of the burst period. Thus, the previous classifier is insufficient for estimating the true time of AF initiation. An alternative classifier to observe, record and track re-entry was thus proposed to estimate AF initiation time.

To model this, we again used the proportion of activated cells to be our `signal', and defined an alternative AF classifier: tracking the proportion of activated cells out of total (non fibrotic) cells, above a non-zero threshold for a period of time. We considered this analogous to clinical monitoring methods such as the ECG, which detect absence of regular peaks (e.g. P waves) for defined periods. Similar methods have been adopted by Manani \cite{Manani:2016}. In the following analysis, we set the non-zero threshold to be $0.5$ and the time period to be $2$ seconds. Using this definition, the classifier operates without perturbing the CA, and the onset time of AF re-entry is a random variable: in different sample paths, the first time the classifier is triggered is random.

\begin{figure}[h!!!]
\begin{center}
\includegraphics[width= 0.38\textwidth]{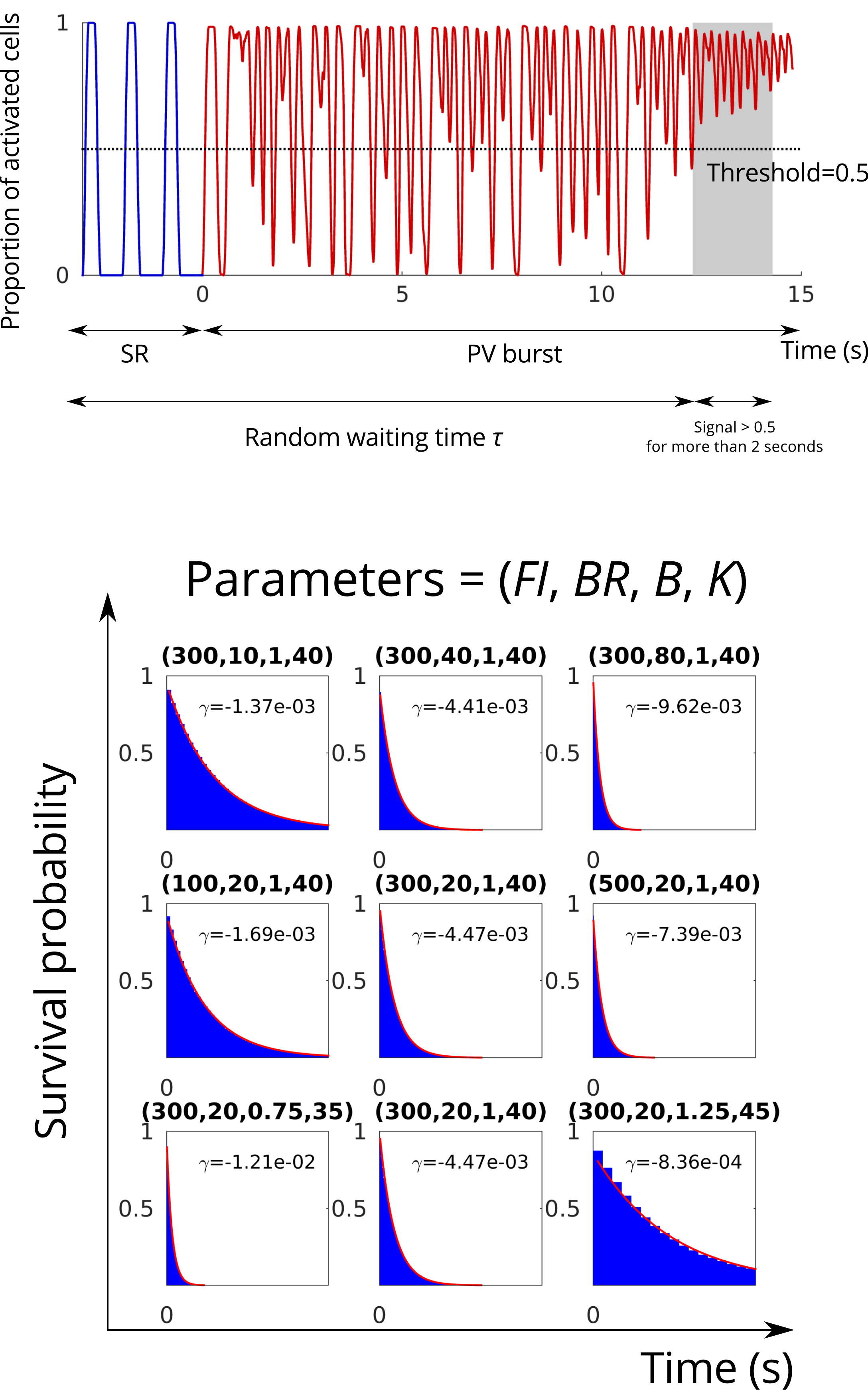}
\end{center}
\caption{\label{fig:Fig4} The second classifier identifies the system as `in AF' if the proportion of active cells is greater than $0.5$ for more than 2 seconds. The onset time of AF is a random time $\tau$, and the cumulative distribution of $\tau$ is plotted in the right panel for selected parameter sets. The data was fitted by an exponential function $\exp \l(-\gamma t\r)$. Note that when we varied BR, the cumulative distribution was monotonically decreasing for any given time. This indicates that the second classifier identifies the transition rate to enter AF is a monotonic increasing function of BR, in contrast to Fig.~3 where AF is mostly induced at the intermediate regime of BR.}
\end{figure}

We refer to the first time the classifier indicates AF as $\tau$. This differs from simulation to simulation, and is random. We simulated $10^5$ samples for selected sets of parameters to compute the cumulative distribution function $s(t)=\mbox{Prob}[\tau > t]$, the probability that the classifier is not activated before time $t$. This monotonically decreasing function quantified the statistics of the random transitions into the first re-entrant episode: the quicker the cumulative distribution function decays, the faster the system transitions to AF on average. The results are presented in Fig.~\ref{fig:Fig4}. The numerical results suggest that the cumulative distribution function is exponential, a signature that the waiting time distribution is also exponential, confirming the simple coarse-grained model with constant rate in Eq.~\eqref{eq:onlyActivationReaction}. 

This analysis suggested a monotonic relation to all parameters, which differs from the first classifier, where burst rate $BR$ showed non-monotonic dependence. We observed that the increased number of PV bursts at high bursting rates raised the proportion of activated cells, triggering the second classifier, but without leading to AF under the definition of the first classifier.   
To test this observation, we performed the following simulation: after the second classifier identified a re-entrant episode, we turned off the PV bursts and evolved the system for another $10$ seconds. We excluded the sample paths which did not exhibit re-entry at the simulation endpoint, following the first classifier. 
For all parameter sets except the high $BR=80$ Hz case, more than $94\%$ of re-entrant episodes identified by the second classifier led to self-perpetuating re-entry. In the $BR=80$ Hz case, only $16\%$ led to re-entry. 

This observation showed the second classifier, albeit realistic in practice, over-estimated the transition rate into AF. 

\noindent{\bf Longer durations of PV bursts suggest existence of spontaneous AF termination dynamics.} 
The above analysis quantitatively estimated timescales for the simulation transitioning into AF. Taking the baseline parameter set $(FI,BR,B,K) = (300,20,1,40)$, both classifiers estimated an $ \approx 5\times 10^{-3} \text{ } 1/\text{s}$ transition rate into AF; in other words, SR is maintained under the influence of PV bursts for $\approx 200$~s. In addition, all sample paths should transit into AF if we waited long enough.

\begin{figure*}[t]
\centering
\includegraphics[width= 0.8\textwidth]{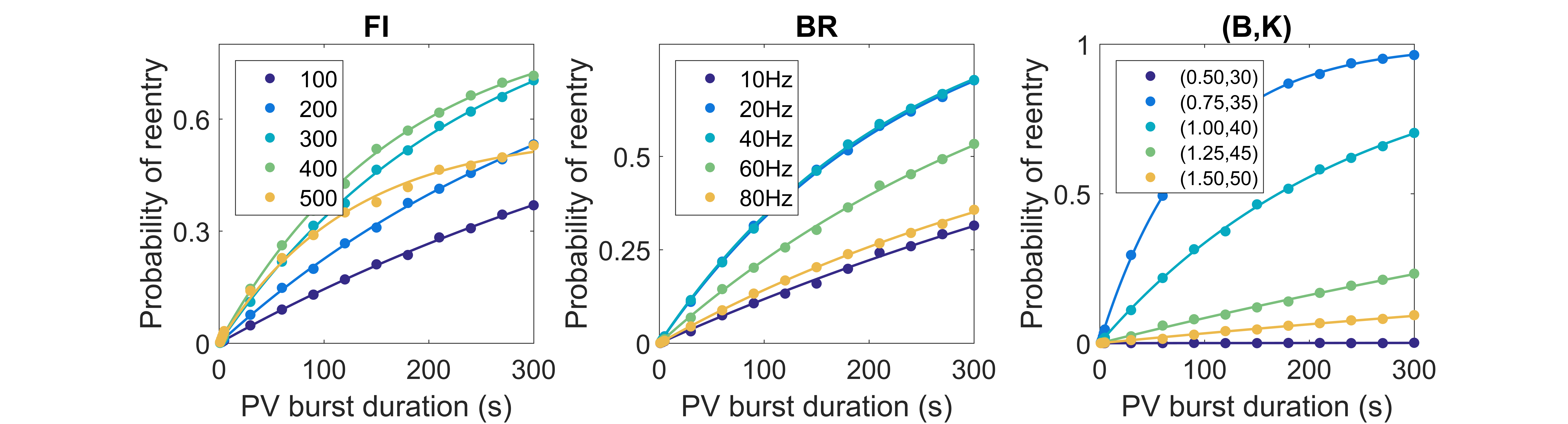}
\caption{\label{fig:Fig5} A parallel analysis of Fig.~\ref{fig:Fig3} to analyse AF initiation rate for longer duration of PV bursts up to 300s. Discrete markers: simulation results; continuous curves: best fits using Eq.~\eqref{eq:2stateTheory}.}
\end{figure*}

To test this assertion, we extended the analysis in Fig.~\ref{fig:Fig3} with a longer PV burst duration. For each of parameter set, we simulated $5000$ sample paths to compute the probability that the sample had transitioned into AF. The results are shown in Fig.~\ref{fig:Fig5}. For some parameter sets, after a long period of PV bursting, the probability did not converge to 1.0 (e.g. for $FI=500$).
In other words, the coarse-grained model Eq.~\eqref{eq:onlyActivationReaction} did not sufficiently capture AF dynamics when the PV burst duration was increased.

We thus generalised the coarse-grained model into a 2-state model with a stochastic initiation and termination of AF under conditions of PV bursting.
\subeq{
\text{SR} \longrightarrow{}& \text{AF} \quad \text{with rate } r_1, \\
\text{AF} \longrightarrow{}& \text{SR} \quad \text{with rate } r_2. 
}{eq:2stateModel}

\begin{figure*}[t]
\centering
\includegraphics[width= 0.8\textwidth]{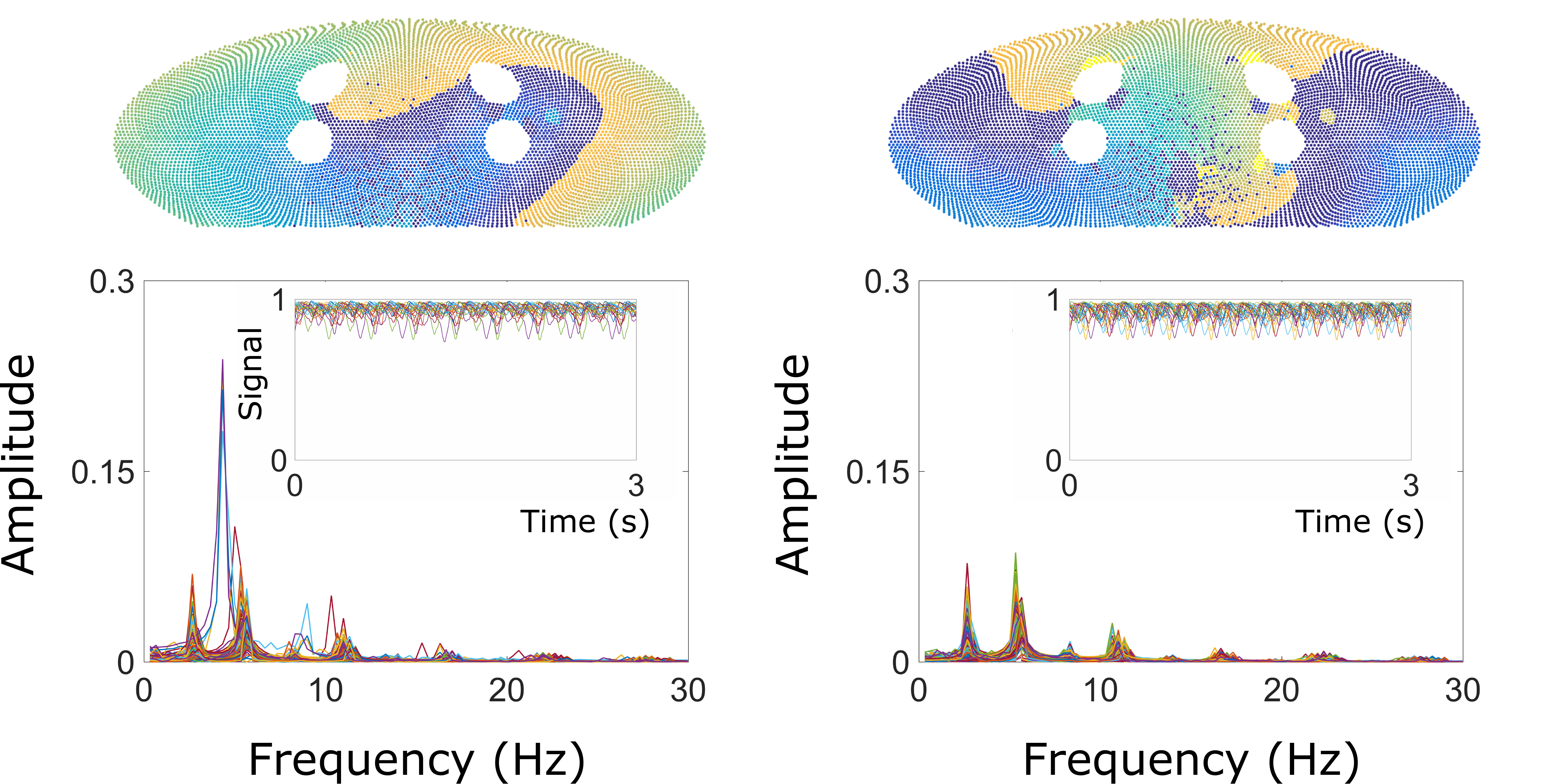}
\caption{\label{fig:Fig6} Fourier analysis of the classifier signal in AF reentry. Both panels used baseline parameter $FI=300$, $BR=20$ Hz, $B=1$, and $K=40$. In the left panel, the Fourier spectrum of 200 sample paths ending with reentry after 5-second PV bursts were overlaid. In the right panel, 200 sample paths with reentry after 150-second PV bursts were plotted. We plotted 20 sample paths on the temporal domain in the insets. Above the figure we present typical snapshots of the visualisation (movies provided in Supplementary Materials), showing a more ``homogeneous'' travelling wave in the left panel, and a more ``fragmented wave'' in the right panel.}
\end{figure*}

Since we start in SR, the initial probability of AF at $t=0$ is always 0 (and ~1 for SR). The temporal behaviour of the probability to be in AF can be calculated using standard methods (see e.g. \cite{Grimmett:2001}), and we find
\eq{
\mbox{Prob}[\text{In AF at time} t] = \frac{r_1}{r_1 + r_2} \l[1 - e^{-\l(r_1 + r_2\r) t}\r].
}{eq:2stateTheory}
A two-parameter fit was performed for each simulated parameter set, and the best fit displayed in Fig.~\ref{fig:Fig6}. Corresponding values of $r_1$ and $r_2$ are reported in Table \ref{tab:parameterset}. The value $r_2$ quantifies the timescale at which stochastic AF initiation is inhibited by constant PV bursts. Comparing the relative values $r_1$ and $r_2$, with high $FI$ or low $BR$, inhibition of AF initiation dominated the process ($r_1<r_2$) and the response of the termination rate to the parameters was also non-trivial.

\noindent{\bf Estimating spontaneous AF initiation and termination times using the model.} To propagate results to our previous model of long timescale AF progression \cite{Chang:2016}, we attempted to project a two-state stochastic model to predict progression of AF at longer timescales. Physiologically, PV bursts occur in acute time periods ($\lesssim 1$ s \cite{Haissaguerre:1998}) rather than chronically. To model this phenomenon, we proposed the following 2-stage and 2-state model:
\subeq{
\text{PV Bursts OFF} \overset{k_1}{\longrightarrow}{}& \text{PV Bursts ON}, \\
\text{PV Bursts ON}  \overset{k_2}{\longrightarrow}{}&  \text{PV Bursts OFF}. 
}{eq:burstingDynamics}
When PV bursts are in ON state,
\subeq{
\quad \text{SR} \overset{r_1}{\longrightarrow}{}& \text{AF} \quad \text{with rate } r_1, \\
\quad \text{AF} \overset{r_2}{\longrightarrow}{}& \text{SR} \quad \text{with rate } r_2.
}{eq:burstingDynamicsII}
Otherwise the state of the system remains in SR/AF respectively. Here $1/k_1$ and $1/k_2$ quantify the average duration of the resting state (no PV bursts) and active state (with PV bursts) respectively. Short trains of bursting means that $k_2 \gg k_1$. Selected parameter regimes were tested (data not shown) and preliminary results showed the coarse-grained model Eq.~\eqref{eq:burstingDynamics} and \eqref{eq:burstingDynamicsII} faithfully projects the progression of the CA model for a range of parameter regimes. However, at much longer timescales $\gg \mathcal{O}\l(1/k_1, 1/k_2\r)$, there were noticeable discrepancies. We investigated these differences in the following section, which suggests existence of higher-order states of AF dynamics.

\noindent{\bf Fourier analysis revealed higher order dynamics of AF.}
Our numerical simulations yielded many sample paths ($\gtrsim 10^4$) which ended in AF re-entry. Fourier analysis was applied to sample paths from the baseline parameter set where re-entry was initiated (corresponding to the 'Observation' phase of the time series shown in Fig.~\ref{fig:Fig3}, upper panel). Results for 200 sample paths are shown in Fig.~\ref{fig:Fig6}. 

For $5$-second PV burst duration (left panel of Fig.~\ref{fig:Fig6}, we visualised sample paths along with the Fourier analysis, and observed that the dominant mode $\approx 5$ Hz corresponds to the period of a single re-entrant wavefront. Sub-dominant half modes $\approx 2.5$ Hz corresponded to the period of points that experienced 2:1 conduction block, e.g. points near one of the PVs which have previously been fast paced. There also existed higher harmonics, to which we did not seek to fit a physiological interpretation. 

There was a noticeable variability in the Fourier spectrum for each sample path. This reflected the stochasticity of the system---including the quenched heterogeneity of RP, fibrosis, and dynamical randomness from PV bursts--- which propagated to the dynamics of re-entry modes. As the speed of the travelling wave is fixed at conduction speed 0.5m/s, the dominant frequency is inversely proportional to the pathlength the wavefront travelled in one cycle. Both the duration of the re-entry and the length of cycle path exhibited $\approx 20\%$ variability.
We also examined the case when PV bursts lasting $150$-seconds were applied (right panel, Fig.~\ref{fig:Fig6}), observing that the variability of the spectrum appeared smaller compared to the $5$-second case. This suggests longer duration of PV bursts tend to drive the system into a stable dynamical mode that is hard to perturb. By comparing visualisations alongside the Fourier spectrum, we also identified that multiple 2:1 conduction blocks formed more frequently, and higher-order rotors were identified. Two representative snapshots are presented in Fig.~\ref{fig:Fig6}.

This analysis shows that even when the model state was classified as `in AF', there can be multiple modes. The follow on question is whether the complexity of an AF episode affects its stability and its likelihood to terminate, either spontaneously or following intervention. For a single re-entrant wavefront, a short PV burst at the right time and location terminated AF (movie on Youtube). This led to an investigation into spontaneous termination of AF in the next section, comparing termination rates for different AF modes, to infer likely mechanisms of termination.

\begin{figure*}[t]
\centering
\includegraphics[width=0.8\textwidth]{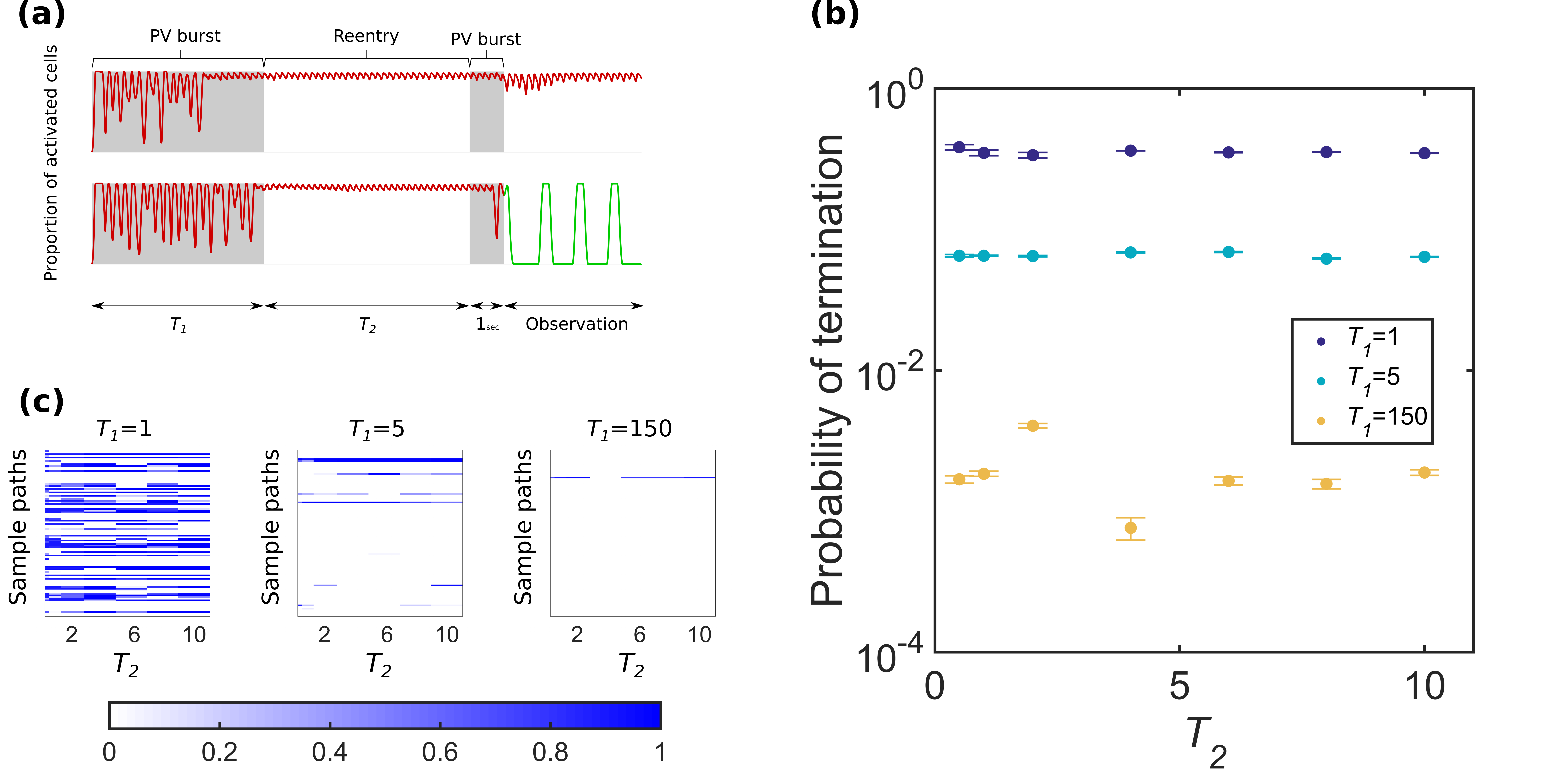}
\caption{\label{fig:Fig7}(a) Schematic diagram of the protocol to probe spontaneous termination. We turned on PV bursts for a duration $T_1$s, to obtain sample paths which initiated AF. For these, a second PV burst of duration $1$s was applied, following a waiting window of duration $T_2$s. Termination probability was measured in the $10$ sec window after the second PV burst. (b) The termination probability across all sample paths depends on $T_1$; in contrast, the value of $T_2$ doe not change the termination probability significantly. (c) We performed 50 trials for $T_2 = [0.5,1,2,4,6,8,10]$, for each sample path in AF after the first PV burst. Probability of AF termination for each sample path (single line), as a function of the sample path and $T_2$, is shown in the heatmap. Strong correlation in the horizontal direction suggests two sub-populations: stable AF which cannot be terminated, and unstable AF which can be terminated, regardless of the $T_2$ value. The results suggests the sub-population of stable AF increases as $T_1$ increases.} 
\end{figure*}

\noindent{\bf Investigation into stochastic AF termination suggests stable and unstable re-entry modes.}
Observation of simulations which generated Fig.~\ref{fig:Fig5} indicated stochastic termination of AF could be a direct result of PV bursts. To test this hypothesis, we randomly collected $500$ sample paths ending in AF in previous experiments and performed $50$ extended simulations on each. Recall that AF was induced by a set of PV bursts over some duration, say $T_1$, in previous experiments. After AF was initiated, we waited a time window $T_2$ without PV bursting, and applied another set of PV bursts (1 sec duration), and observed if re-entry was terminated after the second set of PV bursts had been applied. A schematic diagram of this is shown in Fig.~\ref{fig:Fig7}(a).  

Fig.~\ref{fig:Fig7}(b) shows the termination probability significantly depends on $T_1$. For $T_1=1$ sec, it was very likely to terminate AF, with probability approximately $\gtrsim 0.3$, and it was independent of $T_2$. For $T_1=5$ sec, termination probability was of order $\lesssim 0.1$, and for $T_1=150$ sec the probability went down to order $10^{-3}$. 

This analysis suggests two modes of AF: some re-entrant circuits can be terminated easily by PV bursts, and others cannot. In Fig.~\ref{fig:Fig7}(c), we show probability to termination, ordered by each sample path in the $y$-direction and each waiting window duration $T_2$ in the $x$-direction. A clear alignment in the $x$-direction of either blue or white stripes showed that if a sample path can be terminated, the probability of termination does not critically depend on $T_2$; if the sample path cannot be terminated, most likely, it cannot be terminated for any $T_2$. 
The longer $T_1$ (the duration of the first set of PV bursts to induce re-entry), the smaller the proportion of unstable AF (episodes which can be terminated). Thus the overall probability to terminate AF is orders of magnitude smaller than AF induced by shorter $T_1$. 

Effectively, we hypothesised that activation and termination can be modelled using the multiple-state model:
\subeq{
\text{SR} \rightleftharpoons{}& \text{Unstable AF}, \\
\text{SR} \longrightarrow{}& \text{Stable AF}.
}{eq:multipleState}
Importantly, results suggest that the transition rates are not constant and critically depend on the duration of PV bursts. 
To quantify transition rates, a classifier identifying the signal state must be developed; We aim to develop this in the future. 
Our presented framework can be applied to measure transition rates once a reliable classifier is implemented. 
We remark that the multiple-state system has a ``memory'' for marginal observables (in AF or not) in line with our previously proposed hidden state binary model \cite{Chang:2016}, which can be used to project the progression of AF over long timescales. 

\section*{Discussion}
In this work, we investigated stochastic onset and termination of atrial fibrillation episodes by using a cellular automaton model on a two-dimensional sphere, with a correct topology of the left atrium.  We demonstrated the capability of the model to generate large sets of sample paths to infer statistical properties of AF reentry initiation and termination (up to $\mathcal{O}(10^6)$ sample paths and for duration $\sim \mathcal{O}(10^{4})$ sec). Three potential arrhythmogenic mechanisms were investigated, fibrosis density ($FI$), pulmonary vein bursting rate ($BR$) and refractory period restitution steepness ($B,K$). By probing this parameter space, we investigated the probability of AF onset and termination resulting from PV bursts. 

We found a linear dependence between burst duration and probability of AF initiation for all parameters for short PV burst durations. $FI$ led to a monotonic increase in probability of initiating re-entry, but there was a non-monotonic response for $BR$ and restitution steepness. When PV burst duration was increased, probability of re-entry at simulation endpoint did not increase linearly, such that at high $FI$ and high $BR$, likelihood of re-entry remained constant. This suggests that the sensitivity to the parameters is non-trivial, and indicates the existence of complex dynamics which inhibit AF initiation or possibly terminate re-entrant circuits before they have fully formed. We fitted a 2 state non-linear model to this and estimated initiation and inhibition rates $r_1$ and $r_2$ for given parameter sets.

We implemented a dynamic classifier to estimate time of AF initiation, which overestimated the rate of AF initiation when BR was high compared to our first classifier. Finally, we analysed a subset of the sample paths in AF, and found existence of stable and unstable AF modes. A second set of PV bursts could spontaneously terminate a proportion of induced AF episodes, with termination probability reducing, subject to duration of the first PV bursts. 

We believe this study offers an alternative novel methodology and framework for investigating mechanisms of spontaneous AF, which differ from conventional modelling and experimental studies in its capability for rapid statistical sampling of long timescale episodes. Our conclusions on this study are set out and discussed in the subsequent paragraphs.

\noindent{\bf The Cellular Automaton model is a robust model for investigating AF.} CA models have been superseded by more biophysically detailed models recently \cite{Clayton:2011}, but are still employed, both in standalone theoretical studies and combined with clinical investigations \cite{Bub:2002,Correa:2011,Manani:2016,Spector:2011}. Our approach complements, for example, studies by Manani \cite{Manani:2016}, who similarly used a CA formulation, to investigate effect of time dependent fibrosis on arrhythmia susceptibility. Our model naturally handles variability and uncertainty through its stochastic formulation and the large number of sample paths, and thus permits a systematic investigation within the model framework, whilst accepting the model limitations. A major limitation of the CA model compared to continuum models is its inability to directly model the mechanism of conduction slowing and CV restitution, although most other potential AF mechanisms  \cite{Andrade:2014} may be handled with the CA formulation. 

\noindent{\bf The potential model parameter space is vast.} In this study, we fixed the size and shape of the left atrium, and the size and location of anatomical objects. We did not include the left atrial appendage, and assumed that location of sinus node breakthrough into the left atrium was fixed. Heterogeneity was investigated by randomly varying initial refractory periods, rather than region specific heterogeneity in parts of the left atrium. We recognise that these are all parameters which may vary between individuals, and may significantly impact probability of AF initiation and termination. We chose to focus on biophysical mechanisms rather than anatomical variability, and recognise that additional investigations into the effects of these parameters are important.

\noindent{\bf AF Onset:} Recent studies have investigated mechanisms related to electrical and structural remodelling, highlighting the importance of inter-patient variability. McDowell et. al. \cite{McDowell:2013,McDowell:2015}) found that combinations of fibrosis subtypes were proarrhythmic and that patient specific distribution of fibrosis had a major impact on AF initiation, and anchored wavefronts to specific atrial regions, with other electrophysiological changes not significantly altering this behaviour. Krummen et al \cite{Krummen:2012} reported that steepening AP restitution slope in patients initiated reentry, with the associated computational study identifying specific ionic pathways responsible for restitution steepening. Regional electrical heterogeneity of the atria was investigated by Colman et al \cite{Colman:2013}, who found region-dependent APD heterogeneity in the atrium increased susceptibility to AF onset and maintenance of reentrant circuits. 

Our study has investigated these three mechanisms plus PV firing rate, albeit with a discrete rather than continuous model, and different assumptions and formulations (we did not model fibrosis subtypes or include region-specific RPs for our cells). Our study results differ from the conclusions of these continuum studies, especially regarding the steepening of restitution slope, where we found a non-monotonic relationship between AF onset and restitution steepness not predicted by Krummen et al. There is no general consensus on whether a steep restitution slope is pro- or anti-arrhythmic \cite{Franz:2003}, and our results showed there is a `window' of steepness which maximises probability of AF onset. This was also true for the other parameters, where excessive fibrosis and PV burst rate inhibited increased onset of AF. We comment that a PV burst rate up to 80 $s^{-1}$, whilst representing the number of triggers across all four PVs rather than a single focal source, may appear unphysiological, but it is also possible that many focal PV bursts go undetected.

\noindent{\bf AF Termination:} Clinical studies predominantly investigate how targeted ablations terminate AF, and these have been explored theoretically in a number of studies \cite{Jacquemet:2016}. However, few studies explore spontaneous termination due to the difficulty of capturing such rare events. A few clinical studies have been documented: Ndrepepa \cite{Ndrepepa:2002} referred to generators of fibrillatory activity in the left atrium, and reported that AF termination was polymorphic in its mechanism. Alcaraz \cite{Alcaraz:2008, Alcaraz:2009} analysed the atrial activity of patients during AF and immediately prior to termination, and found the existence of more organised atrial activity (measured by sample entropy) one minute prior to termination, and that the late activity had a significantly lower dominant frequency mean value. Some studies of dominant frequency and harmonics have suggested Fourier analysis as useful predictors of termination \cite{Martins:2014}.

Our study was inconclusive regarding termination. We found that PV bursts are a potential mechanism for terminating as well as initiating AF, and also act to inhibit initiation rate for longer durations of PV bursts. Fourier analysis of the sample paths revealed both stable and unstable modes of AF, but no clear trend was observed. We found however that the longer the period of PV bursting, the smaller the probability that induced AF will be terminated by future PV bursts. This suggests dynamical memory effects exist within the model caused by extended PV burst pacing, which influences the stability and robustness of the induced reentry wavefronts. This agrees with the `AF begets AF concept' \cite{Wijffels:1995}, and recent studies of Uldry et al \cite{Uldry:2012a,Uldry:2009} who reported an increase in AF complexity with duration, and that spontaneous termination mechanisms differed depending on dynamics of AF and its underlying complexity.

In other recent studies, Krogh-Madsen et. al. \cite{KroghMadsen:2012} also found that remodelling maintained AF by shortening atrial wavelength (electrical by shortening APD, structural by slowing conduction), which correlated with increased AF episode duration, with dynamics of reentry differing between types of remodelling. Liberos \cite{Liberos:2016} suggested cell---cell ionic differences as a mechanism of AF termination, by decelerating rentrant activity and increase in rotor tip meandering. Our model did not include electrical remodelling similar to these studies, but our model is well placed to analyse atrial wavelength and track the rotor tips in future studies, to see if similar mechanisms exist within our formulation.  The general consensus is that AF complexity increases over time together with AF episode durations, with size of atria and atrial obstacles thought to play a critical role in termination. Petrutiu \cite{Petrutiu:2007} found that non-terminating episodes exhibited larger dominant frequencies compared to spontaneously terminating episodes, and more abrupt changes in dominant frequency were observed prior to spontaneous termination. An open mechanistic question remains over whether spontaneous termination is preceded by a progressive fusion of wavelets or a simultaneous block of all wavelets in the tissue. We believe our work is well placed to evaluate these questions through the capabilities to run longer time scale simulations.

Our study may additionally complement existing ablation-based termination studies by identifying similar mechanisms or proposing novel therapeutic studies. Rappel \cite{Rappel:2015} demonstrated that ablation caused termination in a heterogeneous domain by creating an excitable gap, dislodging a stable anchored wavefront or by closing critical isthmus channels. Uldry \cite{Uldry:2012} reported 10-20\% success rate when using atrial septal pacing at alternating frequencies to pass the atria. 

\subsection*{Future Work}
Our work in this article focuses on the framework of the stochastic analysis. We acknowledge that CA models are a simplified representation of reality. However it permits large numbers of simulations to obtain probability distributions and probe particular mechanisms. We propose several directions to improve the model:

\noindent{\bf Geometry:} We adopted a simplified quasi-spherical geometry to model the left atrium. Since the dynamical rules of the cellular automaton only involves the neighbourhood relations between nodes, it is straightforward to construct a CA model on any two-dimensional surface embedded in three-space. The difficulty of evenly distributing the nodes may be overcome by using the Archimedean spiral \cite{huttig2008spiral}. It may also be possible to extend this to three dimensions.

\noindent{\bf Directional fibrosis:} In this work we modelled fibrosis by setting nodes to be electrically active, whereas fibrosis may act to promote faster propagation in certain directions within cardiac tissue \cite{Manani:2016}. This could be be achieved in our CA model by assigning weights to neighbouring nodes.

\noindent{\bf Representing interventions:} As the computational cost of a CA model is cheaper than biophysically detailed models, it is an ideal platform to develop and evaluate effects of intervention strategies such as ablation or external pacing. However, as a coarse-grained approach, the CA model is unlikely to capture detailed biochemical or biophysical effects within these, or within other interventions such as pharmacological modification of cell and tissue electrophysiology.

\noindent{\bf Restitution and remodelling:} Restitution changes may not be instantaneous. One way to model restitution with memory would be to replace Eq.~\eqref{eq:restitution} by
\begin{align}
\rm{RP}_{i+1} = {}&\alpha \rm{RP}_{i} \nonumber\\
{}& + \l(1-\alpha \r) \l\{121 \l[1-B\exp(-\rm{DI}/K)\r]\r\}, \label{eq:restitutionModified}
\end{align}
where $\alpha$ measures the strength of the ``memory''. When $\alpha=1$, there exists no restitution, and when $\alpha=0$ it converges to our proposed model \eqref{eq:restitution}. 

Only initial state structural remodelling was investigated in this study. Additional structural and electrical remodelling may be implemented in to the CA framework, both as an initial condition and as a transient process (e.g. with ageing). For example, removing cells (and adding them back) from the domain of excitable cells could model acute scar formation or recovery from ischemia. 

\noindent{\bf Pattern recognition of the reentrant wavefronts:} Our analysis revealed stable and unstable modes of AF. Visualisation of selected sample paths suggested some characteristic differences between these modes: just prior to termination, unstable AF terminated via conduction block through fibrosis regions or pulmonary veins. This often included spontaneous PV bursts at the channel isthmus in a short excitable window. In comparison, stable (did not spontaneously terminate) wavefronts appeared to have more complex pathways of propagation. 

While the Fourier spectrum suggested potential differences, the analysis was inconclusive as there was a large variability over sample paths in a given parameter set. As our classifiers contain only the temporal information, we could additionally use spatial information of the re-entrant wave front to (e.g. rotor tip tracking, local electrogram) to inform our analysis. 

\section*{Supporting Information}
Please visit Github project \href{https://github.com/dblueeye/atrial-fibrillation-cellular-automata}{{\bf/dblueeye/atrial-fibrillation-cellular-automata}} for a working implementation and for movie URLs on Youtube.

\begin{table*}[h]
\centering
    \footnotesize
	\begin{tabular}{ | c | c | c | c| l |}
    \hline
    Parameter $(FI,BR,B,K)$ & Best fit $r_1$ in Fig.~\ref{fig:Fig3} &  Best fit $r_1$ in Fig.~\ref{fig:Fig5} & Best fit $r_2$ in Fig.~\ref{fig:Fig5} \\ \hline\hline 
   $(300,0.05,1,40)$	& $4.395\times 10^{-3}$ & $4.095\times 10^{-3}$ & $6.796\times 10^{-5}$ \\
   $(100,0.05,1,40)$	& $1.603\times 10^{-3}$ & $1.567\times 10^{-3}$ & $9.838\times 10^{-5}$ \\
   $(200,0.05,1,40)$	& $2.646\times 10^{-3}$ & $2.569\times 10^{-3}$ & $9.377\times 10^{-5}$ \\
   $(400,0.05,1,40)$	& $5.558\times 10^{-3}$ & $5.106\times 10^{-3}$ & $7.403\times 10^{-4}$ \\
   $(500,0.05,1,40)$	& $7.451\times 10^{-3}$ & $4.553\times 10^{-3}$ & $3.562\times 10^{-3}$ \\
   $(300,0.025,1,40)$	& $1.395\times 10^{-3}$ & $1.255\times 10^{-3}$ & $2.720\times 10^{-9}$ \\
   $(300,0.1,1,40)$	& $4.069\times 10^{-3}$   & $4.095\times 10^{-3}$ & $6.796\times 10^{-5}$ \\
   $(300,0.15,1,40)$	& $2.403\times 10^{-3}$ & $2.516\times 10^{-3}$ & $1.027\times 10^{-8}$\\
   $(300,0.2,1,40)$	& $1.371\times 10^{-3}$   & $1.580\times 10^{-3}$ & $5.350\times 10^{-4}$ \\
   $(300,0.05,0.5,30)$	&$ 2.451\times 10^{-4}$ &$3.883\times 10^{-6}$ & $1.908\times 10^{-3}$ \\
   $(300,0.05,0.75,35)$	&$1.108\times 10^{-2}$  &$1.125\times 10^{-2}$ & $1.123\times 10^{-5}$\\
   $(300,0.05,1.25,45)$	&$ 9.469\times 10^{-4}$& $8.774\times 10^{-4}$ & $1.463\times 10^{-5}$\\
   $(300,0.05,1.50,50)$	&$4.707\times 10^{-4}$&  $3.417\times 10^{-4}$ & $4.371\times 10^{-4}$\\
\hline
	\end{tabular}
    \caption{\label{tab:parameterset}Numerical results of best fits to the model results using the parameter sets in the first column, for simulations up to 5 seconds using a 2 state linear transition model (2nd column), and up to 300 seconds using a 2 state non-linear model (3rd and 4th columns). $r_1$ is the rate of AF initiation in both models, and $r_2$ is the rate at which initiation is inhibited for the non-linear model. In specific parameter sets, $r_1<r_2$ which suggests a very low rate of AF initiation. } 
\end{table*}

\section*{Competing interests}
We have no competing interests.

\section*{Authors' contributions}
Conceived and designed the experiments: YTL EC RHC. Performed the experiments: YTL EC. Analysed the data: YTL EC. Wrote the paper: all.

\section*{Funding}
We acknowledge support from the UK Engineering and Physical Sciences Research Council (\url{www.epsrc.ac.uk}) grant number EP/K037145/1.

\bibliographystyle{vancouver}

\end{document}